\documentclass[10pt, letterpaper, twocolumns, twoside, final, letter]{IEEEtran}
\usepackage{dsfont}
\usepackage{cite}
\usepackage{graphicx}
\usepackage{amsmath}
\usepackage{amsfonts}
\usepackage{array}
\usepackage{times}
\usepackage{stfloats}
\usepackage{amssymb}
\usepackage{yhmath}
\usepackage{graphics}
\usepackage{textcomp}
\usepackage{amscd}
\usepackage{epsfig}
\usepackage{psfrag}
\usepackage{rotating}
\usepackage{amsmath}
\usepackage{url}
\usepackage{color}
\usepackage{float}

 \newtheorem{thm}{Theorem}

 \newtheorem{Proposition}{Proposition}
\begin{document}
\title{{Mitigating Jamming Attacks Using Energy Harvesting}}
\author{Gada~Rezgui, E. Veronica~Belmega,~\IEEEmembership{Member,~IEEE,} and~Arsenia~Chorti,~\IEEEmembership{Member,~IEEE,} 
\thanks{The authors are with ETIS, UMR 8051, Universit{\'e} Paris Seine, Universit{\'e} Cergy-Pontoise, ENSEA, CNRS, 
France; emails: \{gada.rezgui, belmega, arsenia.chorti\}@ensea.fr.} }

\maketitle

\begin{abstract}

The use of energy harvesting as a counter-jamming measure is investigated on the premise that part of the harmful interference can be harvested to increase the transmit power. We formulate the strategic interaction between a pair of legitimate nodes and a malicious jammer as a zero-sum game. Our analysis demonstrates that the legitimate nodes are able to neutralize the jammer. However, this policy is not necessarily a Nash equilibrium and hence is sub-optimal. Instead, harvesting the jamming interference can lead to relative gains of up to $95\%$, on average, in terms of Shannon capacity, when the jamming interference is high.
\end{abstract}


\section{Introduction}
\label{sec:intro}

In recent years, the simultaneous wireless information and power transfer has gained momentum in the realm of energy harvesting (EH) technologies. In this contribution, we focus on systems employing a time-splitting EH approach \cite{zhou2013, GA,RSV, Ku16}, i.e., during a first phase the receiver collects energy from RF microwave radiation and in a second phase it uses the harvested energy for data transfer. 

An interesting application of such EH approaches arises for wireless systems under jamming attacks. In the past, two main counter-jamming approaches have been commonly considered: direct sequence spread spectrum (DSSS) and frequency hopping spread spectrum (FHSS),
\cite{Wei12,Xiao15}. More recently, the use of multiple antennas has been exploited in \cite{wang2018} against both jamming and eavesdropping. In the first two approaches, the impact of power constrained jammers can be limited by increasing the spectral resources because the optimal jamming strategy is to spread the jamming power over the entire bandwidth; whereas the former requires an increased number of antennas. In the present work, we alternatively explore the possibility of mitigating jamming attacks by using EH \emph{without increasing the spectral or the spatial resources}. So far, there has only been a limited number of contributions in this area \cite{FSL, XCDN, Veronica_ICC17, EVBChorti}. 

In \cite{FSL}, the jamming interference is harvested and exploited in a two-way channel assuming that the jammer's policy is fixed (i.e., the jammer is not strategic). Furthermore, in \cite{XCDN} a cooperative relay wiretap channel is studied, in which the helping nodes harvest energy from the legitimate link and then generate interference to the eavesdropper. On the other hand, in \cite{EVBChorti}, it is shown that EH can be exploited to mitigate jamming attacks in wireless secret key generation (SKG) systems and that it is possible to neutralize the jammer, i.e., to fully compensate its impact on the SKG rate. Building on this idea, the objective of the present work is to investigate whether EH at the legitimate transmitter can be an efficient measure against jamming attacks. We investigate the strategic interaction between a pair of legitimate nodes and a jammer employing as utility function the Shannon capacity.

The main contributions of this paper can be summarized as follows. First, we demonstrate that the jamming attack can be prevented entirely by adjusting the EH duration, i.e., the jammer can be neutralized (or forced to remain silent). This is only possible when the quality of the channel in the harvesting link is higher than in the jamming link. Nevertheless, neutralizing the jammer imposes too stringent restrictions on the EH duration and on the legitimate transmit power and hence is not optimal.

Second, we formulate a zero-sum game between the legitimate users and the jammer and derive the Nash equilibrium (NE) analytically. At the NE, both players transmit at full power, while, the optimal EH duration depends on the system parameters. Interestingly, we show that the NE always outperforms neutralizing the jammer. At the NE, the jamming interference is not fully cancelled but rather exploited, particularly efficient in the high jamming interference regime. 

This work represents a proof of concept of the potential use of EH against jamming attacks, relying on widely used system model assumptions \cite{zhou2013,GA,RSV,Wei12,Xiao15,EVBChorti,Veronica_ICC17}. Demonstration of the proposed EH policy in a real testbed is left as future work, in which the effect of imperfect channel estimation, type I and II jamming detection errors, implementation aspects of EH,  etc., will also be considered.

\section{System model}
\label{sec:system}
\begin{figure}[t]
\setlength{\unitlength}{0.08in} 
\centering 
\begin{picture}(26,11) 
\put(-5.5,6){\framebox(4,4){\small Alice}}
\put(2.75,1){\framebox(4,4){\small Jay}}
\put(11,6){\framebox(4,4){\small Bob}}
\put(-1.5,8){\vector(1,0){12.5}}
\put(4,8.5) {\small $H$}
\put(2.75,3){\vector(-2,1){6}}
\put(-2,3.1) {\small ${G}_{A}$}
\put(6.75,3){\vector(2,1){6}}
\put(10,3.3) {\small ${G}_{B}$}

\put(21,8.5){\small Time sharing EH}
\put(20,4){\line(1,0){13.5} }
\put(20,3.5){\line (0,1){1}}
\put(25,3.5){\line (0,1){1}}
\put(33.5,3.5){\line (0,1){1}}

\put(21.5, 5){\small $\tau T$}
\put(20.5, 1.5){\small EH}
\put(20,-0.5){\small phase}

\put(26.5, 5){\small $(1-\tau) T$}
\put(25.5, 1.5){\small Transmission}
\put(28,-0.5){\small phase}

\end{picture}
\caption{System model and time sharing scheme at Alice's side.}
\label{fig:systemmodel}
\end{figure}
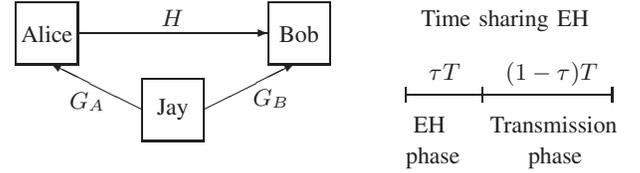

The system model, depicted in Fig. \ref{fig:systemmodel}, comprises three nodes: a legitimate transmitter, Alice, its intended receiver, Bob, and a malicious jammer, Jay. 
The channel coefficients in the links Alice-Bob, Jay-Alice and Jay-Bob are denoted by $H$, $G_A$ and $G_{B}$, respectively and model fading; they are assumed 
to remain constant during each EH and transmission cycle and to change independently from one cycle to the next. We assume that full channel state information is available at all nodes.

When Alice sends a message $X_A$ to Bob, Jay can jam the transmission.   
Bob's observation $Y_B$ can be expressed as:
\begin{eqnarray}
\label{yb}
Y_{B} &=& H X_A + G_{B} X_J+ Z_{B},
\end{eqnarray}
where the message $X_A\sim\mathcal{N}(0,p)$ is drawn from a Gaussian codebook under a short-term power constraint $0\leq p\leq P$. Finally, $Z_{B}\sim\mathcal{N}(0,N_B)$ models the effect of the additive white Gaussian noise (AWGN) in the Alice-Bob link.
We consider that Jay transmits Gaussian jamming signals $X_J\sim\mathcal{N}(0,\gamma)$ and  $0 \leq \gamma\leq \Gamma$ represents the jamming power.

This paper studies whether EH can be exploited to harvest the jamming interference and boost the transmit power.
We assume a time sharing scheme with two phases: the first phase of duration $\tau T$, where $T$ is the symbol period, is dedicated to EH. The second phase of duration $(1-\tau)T$ is dedicated to information transmission. 

Alice's observation during the EH phase is then given by:
\begin{eqnarray}
\label{yaFD}
Y_{A} &=& G_{A} X_J + Z_{A},
\end{eqnarray}
 where $Z_{A}\sim\mathcal{N}(0,N_A)$ models the effect of AWGN noise in the link Jay-Alice.
As commonly assumed \cite{GA}, the harvested energy is proportional to the energy of the received signal:
  \begin{equation}
E =\tau T \zeta  \mathbb{E}\left[|Y_A|^2\right] =\tau T \zeta \left(\gamma |G_{A}|^2 + N_{A}\right), \label{eq:energy}
\end{equation}
where $\zeta \in[0,1]$ is the harvesting efficiency parameter. The average power harvested during the EH phase is used in the information transmission phase and is given as follows:
  \begin{eqnarray}
\label{eq:E}
p^{EH}&=&\frac{\tau}{1-\tau}\zeta \left(\gamma |G_{A}|^2 + N_{A}\right).
\end{eqnarray}
The multiplicative term $\frac{1}{1-\tau}$ in the above expression stems from keeping the energy constant during each transmission phase of duration $(1-\tau) T$ 
\cite{EVBChorti}. The initially available transmit power (aside from the harvested one) is also enhanced to $\frac{p}{1-\tau}$ for the same reason. Under the above assumptions and using standard time sharing arguments \cite{Cover}, the Shannon capacity of the Alice-Bob link is given by
  \begin{equation}
  \label{eq:CEH}
C^{EH}(p, \tau,\gamma) =\frac{(1-\tau)}{2}\log \left( 1 + \frac{\left(\frac{p}{1-\tau} + p^{EH}\right)|H|^2}{\gamma |G_{B}|^2 + N_B}
\right).
 \end{equation}
Note that the multiplicative term $(1-\tau)$ in front of the logarithm represents the reduction of the Shannon capacity due to time sharing.

\section{Jammer Neutralization}
\label{sec:NJ}
In this Section, we first investigate whether it is possible to neutralize the jammer. 
We note that $C^{EH}$ is increasing with the transmit power $p$ for fixed $\tau$ and $\gamma$. On the other hand, $C^{EH}$ is not necessarily decreasing with the jamming power: since the interference is harvested and $p^{EH}$ increases with $\gamma$, the Shannon capacity may even increase with the interfering power $\gamma$ for specific system parameters. Although this may seem counter-intuitive, consider the case in which the interfering link is very poor $|G_B|^2 << 1$ (e.g., Jay is very far from Bob). In this case, the interference at Bob is negligible: $\gamma |G_{B}|^2 + N_B \simeq N_B$ and, hence, $C^{EH}$ increases with $\gamma$ due to $p^{EH}$. We prove this result rigorously by investigating the first-order derivatives of $C^{EH}$.

\begin{Proposition}
\label{prop:threshold}
\textit{
For fixed $p$ and $\tau$, if $\frac{ |G_A|^2}{N_A}> \frac{|G_{B}|^2}{N_B}$, then  $ C^{EH}(p, \tau,\gamma) $ is monotonically increasing w.r.t. $\gamma$ if $p\le p_{th}(\tau)\triangleq \tau K$, with $K\triangleq  \left(\frac{ |G_{A}|^2 N_B}{ |G_{B}|^2}- N_A \right) \zeta$, and it is monotonically decreasing w.r.t. $\gamma$ if  $p > p_{th}(\tau)$. This implies:
\begin{eqnarray}
\arg \max_{\gamma \in [0, \Gamma]} C^{EH}(p,\tau,\gamma)  & = & 0,\text{ if } p\le p_{th}(\tau), \\
\arg \max_{\gamma \in [0, \Gamma]} C^{EH}(p,\tau,\gamma)  & = & \Gamma,\text{ if } p> p_{th}(\tau).
\end{eqnarray}
Otherwise,  $ C^{EH}(p, \tau,\gamma) $ is always  monotonically decreasing w.r.t. $\gamma$ for any fixed $p$ and $\tau$ and
\begin{eqnarray}
\arg \max_{\gamma \in [0, \Gamma]} C^{EH}(p,\tau,\gamma)  & = & \Gamma.
\end{eqnarray}}
\end{Proposition}

Intuitively, if the quality of the harvesting link is higher than that of the jamming link $\frac{ |G_{A}|^2}{N_A}> \frac{|G_{B}|^2}{N_B},$ 
then the legitimate users can neutralize the jammer by tuning the transmit power $p$ and the EH policy $\tau$ such that $p\le p_{th}(\tau)$. This highlights the existence of a power threshold $p_{th}(\tau)$ below which, harvesting the jamming interference in the first phase overcomes the harmful jamming in the second phase. The optimal strategy $(p,\tau)$ that neutralizes the jammer (NJ) is given below.

\begin{thm}  
\label{thm:NJ}
\textit{If $\frac{ |G_A|^2}{N_A}> \frac{|G_{B}|^2}{N_B}$, the strategy that maximizes the capacity while neutralizing the jammer  $(p^{NJ}, \tau^{NJ})$ is given as follows:}

\textit{ a) If $p_{th}^{-1}(P) >1$, then $(p^{NJ},\tau^{NJ}) = ( p_{th}(\hat\tau), \hat\tau)$, 
where $p_{th}^{-1}(p) = \frac{p}{K}$ is the inverse function of $p_{th}(\tau)$.}

\textit{ b) Otherwise, the optimal strategy is
\begin{eqnarray}
   \nonumber
   (p^{NJ},\tau^{NJ}) & =&  \underset{(p,\tau) \in\{(p_1,\tau_1),(p_2,\tau_2)\}}{\arg \max } C^{EH}(p,\tau,0),  \\ \nonumber 
    (p_1,\tau_1)&= &( \min\{p_{th}(\hat\tau),P\}, \min\{\hat\tau,p_{th}^{-1}(P)\}),\\ \nonumber 
    (p_2,\tau_2)&= &(P, \max\{\tilde{\tau},p_{th}^{-1}(P)\}),
\end{eqnarray}
where $\hat \tau\in (0,1)$ and $\tilde{\tau}\in (0,1)$ are the unique solutions of the equations
$$\frac{\partial C^{EH}(p_{th}(\tau),\tau,0)  }{\partial  \tau}=0 \textit{\ and \ } \frac{\partial C^{EH}(P,\tau,0)  }{\partial  \tau}=0$$
respectively, which can be easily computed numerically.\footnote{Finding explicit expressions for $\hat{\tau}$ and $\tilde{\tau}$ is non trivial and involves solving nonlinear equations containing both logarithmic and fractional terms. Instead, we can exploit numerical methods based on iterative one-dimensional search (e.g., {\tt fzero} in MATLAB\textregistered) or methods relying on trust region or Levenberg-Marquardt techniques (e.g., {\tt fsolve} in MATLAB\textregistered).}
}
\end{thm}

\begin{figure}[t]
 \centering
 \includegraphics[width=0.95\columnwidth]{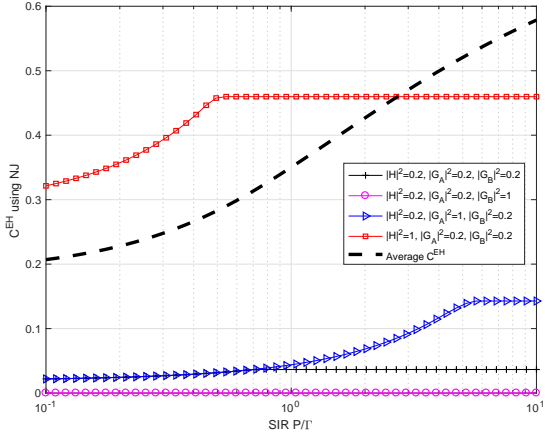}
 \caption{Shannon capacity while neutralizing the jammer as a function of SIR. The jammer cannot be neutralized if the harvesting link is poor. Transmitting at full power $P$ is not always optimal.}
 \label{fig:NJ}
 \end{figure}
Fig. \ref{fig:NJ} illustrates the Shannon capacity obtained while neutralizing the jammer, $C^{EH}(p^{NJ}, \tau^{NJ}, 0)$, as a function of the signal to interference power ratio (SIR) defined as $SIR = P/\Gamma$ in the range $P/\Gamma=-30$ dB to $P/{\Gamma}=10$ dB, for various settings w.r.t. the channels gains for $N_A=-10$ dBm, $N_B=-7$ dBm, $\Gamma=10$ dBm, $\zeta=0.8$. These parameters were chosen to showcase the different regimes in terms of the optimal strategies at the NJ and NE states. Nevertheless, all our remarks and observations are general and remain valid irrespective of the specific choice of the system parameters. 

The capacity $C^{EH}(p^{NJ}, \tau^{NJ}, 0)$ is zero for $|H|^2=0.2$, $|G_{A}|^2=0.2$, $|G_{B}|^2=1$ since the minimum condition on the harvesting link quality is not met and the jammer cannot be neutralized. In the other settings, we can identify two regimes depending on whether $p_{th}^{-1}(P) \leq 1$ or not. When the SIR is low, the system is in case b) of Theorem \ref{thm:NJ} and increasing $P$ increases the feasible set of the optimization problem and, hence, the optimal value of the capacity. At higher SIR, the system shifts to case a), in which the optimal solution that neutralizes the jammer no longer depends on $P$. The dashed black curve depicts the average $C^{EH}(p^{NJ}, \tau^{NJ}, 0)$ over 10,000 random realizations of the channel gains drawn from a standard Gaussian distribution.

We remark that to neutralize the jammer a non-zero EH duration is required, i.e, $\tau^{NJ}>0$, during which little energy can actually be harvested (as the jammer is silent). Also, the transmit power may be required to be below the maximum available power.  
In light of this, we next investigate the optimal strategies of both parties in this competitive interaction.
\section{Optimal Strategies: Game Theoretic Analysis}
\label{sec:GT}

In this Section, we study the adversarial interaction between the legitimate users and the jammer using a non-cooperative game \cite{TiroleFudembergGT}. More specifically, we formulate a two-player zero-sum game defined as the triplet: \\ $\mathcal{G} = \{ \{L,J\}, \{A_{L}, A_{J}\}, C^{EH}(a_{L}, a_{J}) \},$
where $L$ and $J$ denote the opposing players; $A_{L} = [0,P] \times [0,1]$ and $A_{J}=[0, \Gamma]$ denote the possible actions the two players can take; and $C^{EH}(a_L,a_J)$ is both the utility of $L$ and the cost of $J$. The strategy of legitimate users is denoted by $a_{L}=(p,\tau)$ and that of the jammer by $a_{J}=\gamma$. 

A natural solution of the game is the NE, denoted by $(a_{L}^{NE},a_{J}^{NE})$, which is stable to unilateral deviations:
 \begin{eqnarray}
 \nonumber
 C^{EH}(a_{L}^{NE},a_{J}^{NE}) &\ge& C^{EH}(a_{L},a_{J}^{NE}), \ \ \forall a_L\neq a_L^{NE},\\ \nonumber
 C^{EH}(a_{L}^{NE},a_{J}^{NE}) &\le& C^{EH}(a_{L}^{NE},a_{J}), \ \  \forall a_J\neq a_J^{NE}.
 \end{eqnarray}
Neither the legitimate player nor the jammer have any interest in deviating from the NE state knowing that their opponent is playing the NE strategy.

\begin{Proposition}
\label{prop:NJnotNE}
\textit{The NJ state, $(p^{NJ}, \tau^{NJ}, 0)$ in Theorem \ref{thm:NJ}, is not a NE of the game $\mathcal{G}$.}
\end{Proposition}

\textit{Proof:} If $p_{th}^{-1}(P)>1$ then $(p^{NJ},\tau^{NJ})= (p_{th}(\hat\tau),\hat\tau)$, from Theorem \ref{thm:NJ}. This cannot be a NE, because it implies that $p_{th}(\hat\tau)<P$, and since $C^{EH}$ is increasing in $p$ then player $L$ should transmit at maximum power at the NE: $p^{NE}=P$. Otherwise, we have two cases: $ (p^{NJ},\tau^{NJ})= (P, \max\{\tilde{\tau},p_{th}^{-1}(P)\})$ and $ (p^{NJ},\tau^{NJ})= (P, p_{th}^{-1}(P))$. Neither can be NE: since the jammer is silent ($\gamma^{NJ}=0$) and only the noise $N_A$ is harvested, the legitimate user will deviate from $\tau^{NJ}>0$ to $\tau=0$. EH operates as a threat to neutralize the jammer, which results in an inefficient time sharing policy. 

The game's NE is given in the following theorem.
\begin{thm}
\label{thm:NE}
\textit{The NE of the game $\mathcal{G}$ is $(p^{NE}, \tau^{NE},\gamma^{NE}) = (P,\tau^{NE},\Gamma)$, 
where $\tau^{NE}\in\{0,\tau^*\}$ with $\tau^*\in (0,1)$ the unique solution of  
\begin{eqnarray}
\frac{\partial C^{EH}(P,\tau,\Gamma)  }{\partial  \tau}=0
\end{eqnarray}
(which can be easily computed numerically), depending on the system parameters. Moreover, the NE always outperforms the NJ state.}
\end{thm}

\textit{Proof}: The transmit power is maximum $p^{NE}=P$ since $C^{EH}$ is increasing in the transmit power $p$. Then, we prove by \emph{reductio ad absurdum} that at the NE we have: $p_{th}(\tau^{NE})\le p^{NE}$, implying that $\gamma^{NE}=\Gamma$ from Proposition \ref{prop:threshold}. Finding $\tau^{NE}$ reduces to solving the optimization problem:
\begin{eqnarray}
 \tau^{NE} = \arg \max_{\tau \in [0, 1]} \ C^{EH}(P,\tau,\Gamma).
\end{eqnarray}
Based on the first and second order derivatives, $C^{EH}(P,\tau,\Gamma)$ is concave and either decreases ($\tau^{NE}=0$) or it has a unique critical point ($\tau^{NE}=\tau^*$), depending on the parameters. 
To prove that the NE always outperforms the NJ state, we use two ingredients. From Proposition \ref{prop:threshold}, whenever $p=p_{th} (\tau)$, the Shannon capacity is constant w.r.t. $\gamma$ and, hence, $C^{EH} (p^{NJ}, \tau^{NJ},0) = C^{EH}(p^{NJ},\tau^{NJ}, \Gamma)$. From the NE definition and knowing that $p^{NE} =P$ and $\gamma^{NE}=\Gamma$:
\begin{eqnarray}
(P, \tau^{NE})= \arg\max_{p,\tau} \ C^{EH} (p,\tau, \Gamma).
\end{eqnarray}
These two facts yield that the NE outperforms the NJ state: $C^{EH} (p^{NJ}, \tau^{NJ},0)  \leq C^{EH}(P, \tau^{NE}, \Gamma)$. 

\begin{figure}[t]
 \centering
 \includegraphics[width=0.95\columnwidth]{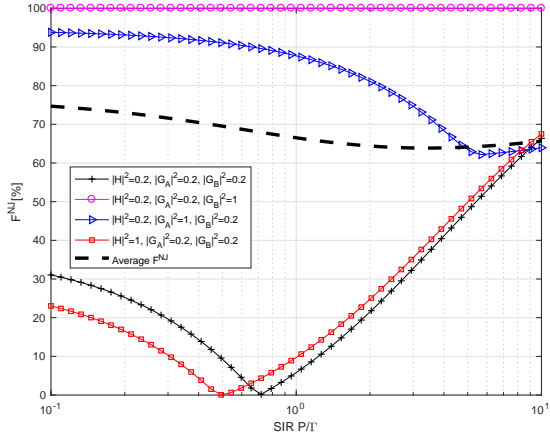}
 \caption{NE vs. NJ efficiency: $F^{NJ}$ in \eqref{eq:FNJ} as a function of SIR. The NE always outperforms the NJ state. At low SIR, exploiting the dominant jamming interference is more beneficial than silencing the jammer. At high SIR, neutralizing the jammer via EH is inefficient.}
 \label{fig:NEvsNJ}
 \end{figure}
 
 \begin{figure}[t]
 \vspace{-0.2in}
\begin{center}
 \includegraphics [width=0.95\columnwidth]{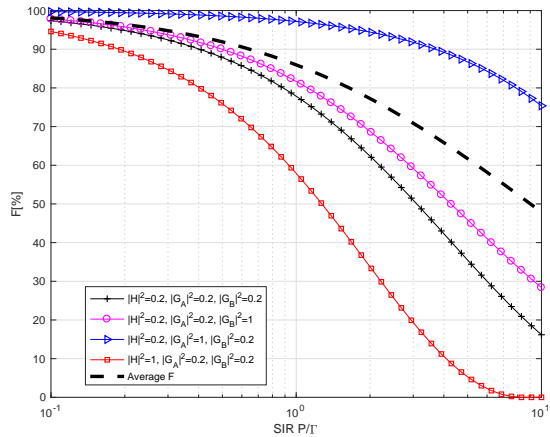}
 \caption{EH efficiency: $F$ in \eqref{eq:F} as a function of SIR. EH is particularly beneficial at low SIR when the dominant jamming interference can be exploited. At high SIR, the interference becomes negligible and EH is less useful.}
 \label{fig:NEvsnoEH}
 \end{center}
 \vspace{-0.2in}
 \end{figure}

In Fig. \ref{fig:NEvsNJ}, we compare the capacity at the NE, $C^{EH}(P,\tau^{NE}, \Gamma)$, with the capacity when neutralizing the jammer, $C^{EH}(p^{NJ},\tau^{NJ},0)$, normalized to the former by illustrating 
\begin{eqnarray}
\label{eq:FNJ}
F^{NJ} \triangleq\frac{C^{EH}(P,\tau^{NE},\Gamma) - C^{EH}(p^{NJ},\tau^{NJ}, 0) }{C^{EH}(P,\tau^{NE},\Gamma)}.
\end{eqnarray}
The simulation setting is identical to Fig. \ref{fig:NJ}. We remark that the NE always outperforms the NJ policy, which is consistent to our analysis. The intuition is that, when neutralizing the jammer, Alice does not necessarily transmit at maximum power $P$ and has to spend a minimum proportion of time $\tau^{NJ}>0$ for EH, as a threat to force the jammer to remain silent, even though no energy can actually be harvested during this time. Only in two of the four specific channel settings, the capacity at the NJ state equals the NE capacity and only at specific SIR values (the $0\%$ minimum points). Surprisingly, using EH to neutralize the jammer is not beneficial in most cases. Instead, the legitimate users should harvest the jamming interference and use it for information transmission. The $100\%$ relative gain curve corresponds to the case in which the jammer cannot be neutralized because of the poor quality of the harvesting link. The dashed black curve represents the average efficiency of the NE w.r.t. the NJ state and shows a relative gain between $64\%-75\%$.

In the same setting, in Fig. \ref{fig:NEvsnoEH} we evaluate the efficiency of EH as a measure against a strategic jammer and compare the capacity at the NE, $C^{EH}(P,\tau^{NE}, \Gamma)$, with $C^{EH}(P,0,\Gamma)$, the capacity in absence of EH  capability normalized to the former, by analyzing 
\begin{eqnarray}
\label{eq:F}
F \triangleq \frac{C^{EH}(P,\tau^{NE},\Gamma) - C^{EH}(P,0,\Gamma) }{C^{EH}(P,\tau^{NE},\Gamma)}.
\end{eqnarray}
At low SIR, the relative EH gain in terms of capacity is very high, approaching $100\%$ when the gain of the harvesting link is high. The jamming interference is dominant and exploiting it for useful communication is very beneficial. At high SIR, the gain from EH decreases progressively towards zero since the jamming interference that can be harvested becomes negligible. The potential of using EH as an anti-jamming measure is demonstrated by the substantial relative gains in terms of Shannon capacity that can on average reach $95\%$ in the low SIR regime.

\section{Conclusion}
\label{conclu}
In this paper, we showed that EH can be exploited to efficiently mitigate jamming attacks.
We proved that a jammer can be completely neutralized by appropriately tuning the transmit power and the EH duration. However, this restricts the transmit power of the legitimate user and requires a minimum harvesting duration, during which little energy can actually be harvested (as the jammer is forced to remain silent). Therefore, neutralizing the jammer is not necessarily optimal. Employing a zero-sum game formulation, we showed that at the NE both players should transmit at full power and the optimal EH duration depends on the system parameters. 
Our simulation results show that  EH can offer substantial gains in terms of capacity. 
At low SIR, the average gains can reach $95\%$, showcasing the high potential of EH as an efficient counter-jamming measure.

\bibliographystyle{IEEEtran}
\normalsize
\bibliography{belmega_WCL2018-0885_biblio}

\end{document}